\documentclass[draft,preprint,aps,%superscriptaddress,
showpacs%,nofootinbib,eqsecnum,preprintnumbers,twocolumn,prl
]{revtex4}

\usepackage{epsfig}

\usepackage{graphicx}% Include figure files
\usepackage{bm}% bold math

\input{epsf}

\newcommand{\be}{\begin{equation}}
\newcommand{\ee}{\end{equation}}
\newcommand{\bea}{\begin{eqnarray}}
\newcommand{\eea}{\end{eqnarray}}
\newcommand{\nn}{\nonumber \\}

\newcommand{\Tr}{\mbox{Tr}}
\newcommand{\tr}{\mbox{tr}}

\newcommand{\psla}{\mbox{$\,\partial\!\!\!/\,$}}

\begin{document}

\preprint{Guchi-TP-020}
\date{\today%
}
\title{Induced Gravity from Theory Space}

\author{Nahomi Kan}
\email{b1834@sty.cc.yamaguchi-u.ac.jp}
\affiliation{Graduate School of Science and Engineering, Yamaguchi University, 
Yoshida, Yamaguchi-shi, Yamaguchi 753-8512, Japan}

\author{Kiyoshi Shiraishi}
\email{shiraish@yamaguchi-u.ac.jp}
\affiliation{Faculty of Science, Yamaguchi University,
Yoshida, Yamaguchi-shi, Yamaguchi 753-8512, Japan}

\begin{abstract}
The mass spectrum of a model constructed in a theory space is expressed 
by eigenvalues of the Laplacian on the graph structure of the theory space. 
The nature of the one-loop UV divergence in the vacuum energy is then 
controlled only by the degree matrix of the graph.
Using these facts, we can construct models of induced gravity which 
do not suffer from divergences at the one-loop level.
\end{abstract}



\maketitle

%%%%%%%%%%%%%%%%%%%%%%%%%%%%%%%%%%%%%%%%%%%%%%%%%%%%%%%%%%%%%%%%%%%%%%
%%%%%%%%%%%%%%%%%%%%%%%%%%%%%%%%%%%%%%%%%%%%%%%%%%%%%%%%%%%%%%%%%%%%%%
%%%%%%%%%%%%%%%%%%%%%%%%%%%%%%%%%%%%%%%%%%%%%%%%%%%%%%%%%%%%%%%%%%%%%%

%%%%%%%%%%%%%%%%%%%%%%%%%%%%%%%%%%%%%%%%%%%%%%%%%%%%%%%%%%%%%%%%%%%%%%
\section{Introduction}
%%%%%%%%%%%%%%%%%%%%%%%%%%%%%%%%%%%%%%%%%%%%%%%%%%%%%%%%%%%%%%%%%%%%%%

The idea of dimensional deconstruction~\cite{ACG} has appeared recently.
A number of copies of a four-dimensional `theory' as well as
a new set of fields linking pairs of these 
`theories' are considered. 
Then the resulting whole theory given by the `theory space' may be equivalent to a 
higher-dimensional theory with discretized extra dimensions.

The simplest model with the Abelian symmetry is studied by Hill and
Leibovich~\cite{HL}. The lagrangian density for vector fields is written as
\be
{\cal L}_A=\frac{1}{g^2}\sum_{k=1}^{N}
\left[-\frac{1}{4}F_{k}^{\mu\nu}F_{k~\mu\nu}-
(D^{\mu}U_{k})^{\dagger}D_{\mu}U_{k}\right]\, ,
\label{lag}
\ee
where $g$ is a gauge coupling,
$F_{k}^{\mu\nu}=\partial^{\mu}{A}_{k}^{\nu}-
\partial^{\nu}{A}_{k}^{\mu}$
and $\mu, \nu=0, 1, 2, 3$. We should read $A^{\mu}_{N+1}=A^{\mu}_{1}$, {\it etc}. The
link fields
$U_{k}$ are transformed as
$U_{k}\rightarrow W_{k}U_{k}W_{k+1}^*$ ($|W_{k}|=1$ for any $k$).
The covariant derivatives are defined as
$D^{\mu}U_{k}=\partial^{\mu}U_{k}-i{A}_{k}^{\mu}U_{k}+
iU_{k}{A}^{\mu}_{k+1}$.

Now we assume that the absolute value of each link field $|U_{k}|$ has a common
value 
$f/\sqrt{2}$. Then the part of the lagrangian for the link fields
gives a mass term for the vector fields in the background field:
For example, if $N=5$, the $(mass)^2$ matrix takes the form%
\footnote{As explained in \cite{HL}, the massive modes absorb the modes of the link
field, and the zero-mode of the link field survives.}
\be
f^2\left(
\begin{array}{ccccc}
2 & -1 & 0 &0&-1\\
-1 & 2 & -1 &0&0\\
0 & -1 & 2 &-1&0\\
0& 0 & -1 &2&-1\\
-1 & 0 &0& -1 &2
\end{array}\right)\, .
\label{cm}
\ee

As seen later, up to the dimensional coefficient $f^2$,
this matrix is identified with the Laplacian matrix for the graph $C_N$,
the cycle graph with $N$ vertices.
We find, indeed, any theory space can be associated with the graph if the link fields
have a common value.

Similarly, field theory on the space as a discrete set of points has been studied
by many authors~\cite{ma}.
We use the `graph theory space' as an extra space,
so we do not worry about the naive continuous limit;
but we study the UV divergent behavior of the four-dimensional one-loop effective
lagrangian.

In Sec.~\ref{sec:2}, we study the Laplacian of the
graph theory space.
The one-loop effective lagrangian density is discussed in Sec.~\ref{sec:3}.
We construct models of induced gravity Sec.~\ref{sec:4}, by using several graphs. We
close with Sec.~\ref{sec:5}, where summary and prospects are given.

%%%%%%%%%%%%%%%%%%%%%%%%%%%%%%%%%%%%%%%%%%%%%%%%%%%%%%%%%%%%%%%%%%%%%%
\section{Laplacian for graph theory space}
\label{sec:2}
%%%%%%%%%%%%%%%%%%%%%%%%%%%%%%%%%%%%%%%%%%%%%%%%%%%%%%%%%%%%%%%%%%%%%%

In this section, after a brief description of graph theory%
%at an introductory level
~\cite{graphtheory},
the Laplacian for a graph is introduced.\footnote{
More discussion on the mathematical description will be found in~\cite{fw}.}
Suppose $G=(V,E)$ is a simple graph,
where $V$ is the set of vertices while $E$ is the set of edges
connecting two vertices.
Here `simple' means that the graph does not include multiple edges and self-loops.
The order of $G$, denoted by $p$, is the number of vertices in the graph
and the size of $G$, denoted by $q$, is the number of edges in the graph.  

A pair of vertices $v_i$ and $v_j$ are said to be adjacent
if there exists an edge $e\in E$ such that $e=\{v_i,v_j\}$.
The adjacency matrix $A(G)=(a_{ij})_{v_i,v_j\in V}$ is defined as
\be
a_{ij}=\left\{
\begin{array}{ccl}
1&:&{\rm if}~v_i~{\rm is~adjacent~to}~v_j\\
0&:&{\rm otherwise}
\end{array}
\right.\, .
\ee

The degree of a vertex $deg(v)$ is the number of edges directly connected to $v$.
The (diagonal) degree matrix $D(G)=(d_{ij})_{v_i,v_j\in V}$ is defined as
\be
d_{ij}=\left\{
\begin{array}{ccl}
deg(v_i)&:&{\rm if}~v_i=v_j\\
0&:&{\rm otherwise}
\end{array}
\right.\, .
\ee

Then we define the Laplacian matrix of $G$ as 
\be
\Delta(G)=D(G)-A(G)\, .
\ee

\begin{figure}[htb]
\centering
\mbox{\epsfbox{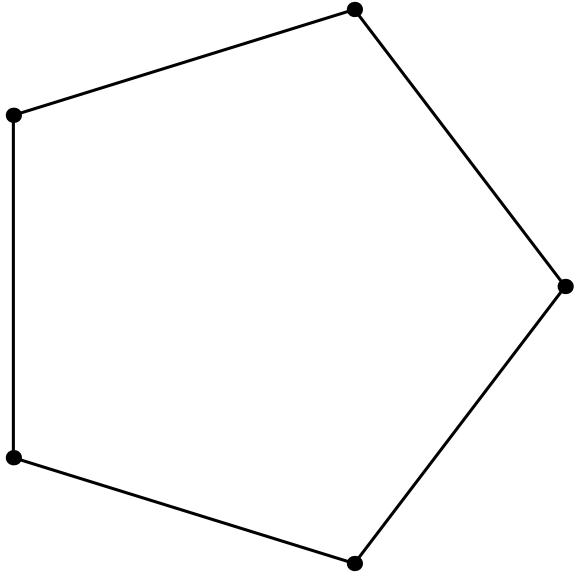}}\\
%\mbox{(a)}\\
%\mbox{\epsfbox{scalar2.eps}}\\
%\mbox{(b)}\\
\bigskip
\caption{$C_5$. $p=q=5$ for this graph.}
\label{fig1}
\end{figure}

The $(mass)^2$ matrix for vector fields in the Hill-Leibovich
model~\cite{HL}((\ref{cm}) for
$N=5$) is written as
$f^2\Delta(C_N)$, where $C_N$ is the cycle graph with $N$ vertices (FIG.~\ref{fig1}),
because $D(C_N)=diag.~(2,2,\cdots,2)$ and only non-zero elements of $A(C_N)$ are 
$a_{i,i\pm 1}$ (where the suffices should be read in modulo $N$).

%%%%%%%%%%%%%%%%%%%%%%%%%%%%%%%%%%%%%%%%%%%%%%%%%%%%%%%%%%%%%%%%%%%%%%
%%%FIGURES 1
%%%%%%%%%%%%%%%%%%%%%%%%%%%%%%%%%%%%%%%%%%%%%%%%%%%%%%%%%%%%%%%%%%%%%%
%%%%%%%%%%%%%%%%%%%%%%%%%%%%%%%%%%%%%%%%%%%%%%%%%%%%%%%%%%%%%%%%%%%%%%
%%%FIGURES 1
%%%%%%%%%%%%%%%%%%%%%%%%%%%%%%%%%%%%%%%%%%%%%%%%%%%%%%%%%%%%%%%%%%%%%%

The lagrangian for the vector fields can be naturally constructed
by putting the $D_{\mu}UD^{\mu}U$ terms on the corresponding edges.
We must put plaquette type terms to give masses to the neutral scalar fields
originated from the link fields~\cite{Lane}. In this paper, we concentrate the case
with
$p=q$  (which holds for $G=C_p$). Then we have only one massless scalar field in
addition to the vector fields.

On the other hand,
for elementary scalar fields, it is easy to write the lagrangian such as
$f^2\phi_i\Delta_{ij}\phi_j$, as long as gauge symmetry is omitted.
The link fields can be attached according to the symmetry of each theory on the
vertex, such as $\phi_i\phi_j\Rightarrow |U_{\{i,j\}}|^{-1}\phi_iU_{\{i,j\}}\phi_j$,
where the gauge transformations $\phi_k\rightarrow W_k\phi_k$ and
$U_{\{i,j\}}\rightarrow W_iU_{\{i,j\}}W_{j}^{\dagger}$ are imposed.

For the Dirac fields, Hill and Leibovich considered the
Wilson fermion on $C_N$~\cite{HL}. Here we suppose the simple case with
$U_{k}={f}/{\sqrt{2}}$. Then the lagrangian for the Wilson fermion
with a unit hopping parameter takes the form
\bea
{\cal L}_{\psi}&=&
\sum_{k=1}^{N}
\left[i\bar{{\Psi}}_{kL}\psla{\Psi}_{kL}+
i\bar{{\Psi}}_{kR}\psla{\Psi}_{kR}\right]\nn
&+&\sum_{k=1}^{N}
f\left[\bar{{\Psi}}_{kL}\left({\Psi}_{k R}
-{\Psi}_{k+1,R}\right)
+{\rm h.c.}\right]\, .
\eea
If we wish to generalize the lagrangian to the case with an
arbitrary graph of $p=q$,
we first introduce the incidence matrix $B(G)=(b_{ij})_{v_i\in V,e_j\in E}$ defined as
\be
b_{ij}=\left\{
\begin{array}{ccl}
1&:&{\rm if~the~edge}~e_j~{\rm is~connected~to~the~vertex}~v_i\\
0&:&{\rm otherwise}
\end{array}
\right.\, .
\ee
By rearranging the label of vertices, we make the diagonal elements of $B$ unity.
Let $E$ be the matrix whose elements are equal to $B$ but
off-diagonal elements change the sign. 
$E$ is often quoted as the incidence matrix for the directed graph in some literature.
The following relation is known:
\be
E E^T=\Delta(G)\, .
\label{pq}
\ee
We can write the mass matrix part of the lagrangian in general, as
\be
f\left[\bar{{\Psi}}_{iL}E^T_{ij}{\Psi}_{j R}
+{\rm h.c.}\right]\, .
\ee
The inclusion of the link variable is possible;
we will see that later in the certain model and more rigorous treatment will be found
in~\cite{fw}.

Now the Dirac fermions on $G$ have the $(mass)^2$ matrix $f^2\Delta(G)$
as in the case for the bosonic fields.
Only two comments are in order here:
First, for models deconstructing higher dimensions, we consider some 
Cartesian product of graphs.
Second, by construction, we find that one chirality of fermions are {\it naturally}
associated with edges (not with vertices), and Eq.~(\ref{pq}) holds even for $p\ne
q$.%
\footnote{Eventually, Arkani-Hamed {\it et al.} argued that the link field originates
from the condensation of fermions. We do not pursue something about this scenario
here.}

Many interesting results on the graph theory have been found in the mathematical
literature. 
The spectrum of the Laplacian for graphs
is studied in the mathematics community.%
\footnote{See \cite{Chung} and references therein. Note however that
this book mainly treat the spectrum of another but related Laplacian.}
It is straightforward
to figure out the mass spectrum of the field theory in theory space
of a general graph by using the mathematical results.%
\footnote{In general Kaluza-Klein theories, the bound for the first excited mass level
is discussed in~\cite{specR}. }
In the rest of the present paper, however, we will concentrate on the study of
one-loop UV divergence of the model in the graph theory space.
We leave the estimation of the spectrum and the physical discussion of the 
observability of excited modes for future work~\cite{fw}.

%%%%%%%%%%%%%%%%%%%%%%%%%%%%%%%%%%%%%%%%%%%%%%%%%%%%%%%%%%%%%%%%%%%%%%
\section{one-loop effective lagrangian}
\label{sec:3}
%%%%%%%%%%%%%%%%%%%%%%%%%%%%%%%%%%%%%%%%%%%%%%%%%%%%%%%%%%%%%%%%%%%%%%

For a gaussian scalar field in the graph theory space $G$,
the partition function is defined as
\be
Z=\int {\cal D}\phi \exp\left[-\frac{1}{2}\phi
\left(-\nabla^2+\Delta(G)\right)\phi\right]\, ,
\ee
where we omitted the indices of scalars as well as the integration on the spacetime
($\int d^4x$), and $\nabla^2$ is the usual laplacian for the scalar field.
Further, we set the factor of scale $f$ to unity for simplicity.
Performing the gaussian integral, we obtain
\bea
Z\propto \left[\det\left(-\nabla^2+\Delta(G)\right)\right]^{-1/2}
&=&\exp\left[-\frac{1}{2}\ln\det\left(-\nabla^2+\Delta(G)\right)\right]\nn
&=&\exp\left[-\frac{1}{2}\tr\ln\left(-\nabla^2+\Delta(G)\right)\right]\, .
\eea
The argument of the exponential in the right hand side is the one-loop
effective action. Using the Schwinger parameter, one can find
\be
\ln\left(-\nabla^2+\Delta(G)\right)
=-\int_{1/\Lambda^2}^{\infty}\frac{dt}{t}\exp\left[
-\left(-\nabla^2+\Delta(G)\right)t\right]\, ,
\ee
where $\Lambda$ is the UV cutoff scale.

In the flat $d$-dimensional spacetime, the part including the $d$-dimensional
laplacian becomes
\be
\tr\exp\left[-\left(-\nabla^2\right)t\right]=\int\frac{d^dp}{(2\pi)^d}
\exp\left[-p^2t\right]=
\frac{1}{(4\pi)^{d/2}}t^{-d/2}\, .
\label{14}
\ee

On the other hand, we find
\be
\Tr\exp\left[-\Delta(G) t\right]=
p-(\Tr D)\,t+\frac{1}{2}\left[\Tr D^2+\Tr D\right]t^2-+\cdots\, ,
\label{15}
\ee
since $\Tr A^2=\Tr D$ holds for an arbitrary simple graph.

From (\ref{14}) and (\ref{15}), we find that the UV divergent terms
depend only on the degree matrix $D$ (and the order $p$) for $d=4$.
In particular, the theory on the graph with the same degree matrix,
even if its adjacency matrix is different,
has the same one-loop divergences in vacuum.%
\footnote{In general, of course, the same divergences can be found in the
case with the common values of $\Tr D$ and $\Tr D^2$ respectively and $D$ itself is
not necessarily identical.
 For example,
$D(G_1)=diag.~(1,1,2,2,2,2,2,4)$ and
$D(G_2)=diag.~(1,1,1,2,2,3,3,3)$ satisfy the condition
$\Tr D(G_1)=\Tr D(G_2)$ and $\Tr D^2(G_1)=\Tr D^2(G_2)$ (and $p=q$).}

We consider the case that the appropriate link fields are associated
and a zero-mode field $\delta$ is survived as
$U\propto \exp (i\delta)$.
The corresponding adjacency matrix $A$ has the non-zero element $a_{ij}$ only for the
adjacent  pair $v_i,v_j$, but in this case the value is $\exp (\pm i\delta)$ instead
of unity. In addition, $A$ is an hermite matrix.
Even in this case, $\Tr A^2=\Tr D$ holds, therefore the divergent term is unchanged;
in other words, the one-loop effective potential for the zero-mode of the link field
can be defined unambiguously up to the constant.

Thus the zero-mode field of the link variable acquires mass by the one-loop quantum
effect. We can consider the phase variable as a constant instead of a dynamical
variable. Then the mechanism of generation of mass can be regarded as a
generalization of the Scherk-Schwarz mechanism~\cite{SS}, since the theory on the
graph
$C_N$ in the limit of $N\rightarrow\infty$ corresponds to its compactification on
$S^1$.

If some non-Abelian gauge symmetry is introduced, a symmetry breaking mechanism
becomes possible in the theory on a graph, just as in the case of the Hosotani
mechanism~\cite{Hosotani}.
This generalization of the Hosotani mechanism will be studied in future work. 

%%%%%%%%%%%%%%%%%%%%%%%%%%%%%%%%%%%%%%%%%%%%%%%%%%%%%%%%%%%%%%%%%%%%%%
\section{one-loop finite Newton constant from graph theory space}
\label{sec:4}
%%%%%%%%%%%%%%%%%%%%%%%%%%%%%%%%%%%%%%%%%%%%%%%%%%%%%%%%%%%%%%%%%%%%%%

If the spacetime is curved, the scalar curvature arises in the effective lagrangian.
Then we have the Einstein-Hilbert action as a quantum effect.
Such an `induced gravity' scenario has been studied since
ninteen-sixties~\cite{Sakharov}. For a recent review, see~\cite{Visser}.

In the curved space, we find
\be
\tr\exp\left[-\left(-\nabla^2\right)t\right]=\int\frac{d^dp}{(2\pi)^d}
\exp\left[-p^2t\right]=
\frac{\sqrt{g}}{(4\pi)^{d/2}}t^{-d/2}(a_0+a_1 t+\cdots)\, ,
\ee
where $a_0$, $a_1$,\ldots depends on the background fields.
For a minimal scalar field, $a_0=1$ and $a_1=(1/6)R$, where $R$ is the scalar
curvature.
For a Dirac spinor field, $a_0=2^{[d/2]}$ and $a_1=-2^{[d/2]}(1/12)R$.
At the same time, we also remember that the fermion loop demands an extra minus sign.
For a massive vector field, $a_0=d-1$ and $a_1=[-1+(d-1)/6]R$.
Then the effective lagrangian takes the form
\be
-V_0+\frac{1}{16\pi G_N}R+\cdots\, ,
\ee
where $V_0$ stands for the vacuum energy density and $G_N$ is the induced Newton
constant.

If we want to calculate the Newton constant and the cosmological constant
(which is often defined as $8\pi G_N V_0$)
unambiguously at one-loop level,
we should consider the cancellation of the divergence.
Because the one-loop contribution of bosonic and fermionic fields have different
signs, the cancellation is possible if we choose the matter content of the field
theory. The complete cancellation of divergence can be expected if the theory has
supersymmetry. In such a case, however, it is difficult to obtain a finite value of
the Newton constant.

In the last section, we saw that the one-loop divergences are governed by the degree
matrix of the graph in the theory on the graph.
Even if the individual theory on the vertex is identical, the different graphs
for the different fields are possible. We can choose the graphs and matter content to
cancel the one-loop divergences.%
\footnote{The induced gravity models were considered in the context of the
Kaluza-Klein theories~\cite{KKIG}. The higher-dimensional divergences are regularized
but the treatment of them is unclear in a certain sense.}

For example, we will investigate a simple case.
Consider a disconnected graph $G=G_1\cup G_2\cup\cdots\cup G_s$ and $G_i\cap
G_j=\phi$ (the empty set) for any $i, j=1,2,\cdots s$. Then let
$G_i=C_{p_i}$, where $p_1+p_2+\cdots+p_s=p$. Any choice of $p_i$ yields
the same degree matrix $diag.~(2,2,\cdots, 2)$.  Since $p'$ is equal to or
more than three for a $C_{p'}$, the total order $p$ is equal to or more than six. 
Thus the appropriate choice of matter content bring about the cancellation of
divergences among the different graphs.

Further in this section, we exhibit minimal models with $p=6$ explicitly.

In the first case,
we consider a vector field theory on $G_V=C_6$, which is described by the lagrangian
(\ref{lag}). We take the zero-mode of the link field into account here, so
we set $U=(f/\sqrt{2})\exp(i\delta)$ for all of the link fields associated with edges.
A Dirac field theory is assumed on the same $G_F=G_V=C_6$. We try to consider the
coupling between the Dirac field and the link field. According to \cite{HL}, the
kinetic term on each vertex and the `mass term'
\be
f\sum_k\left[\bar{{\Psi}}_{kL}\left({\Psi}_{k R}
-e^{i\delta}{\Psi}_{k+1,R}\right)
+{\rm h.c.}\right]\, ,
\ee
is required in the lagrangian if only the link zero-mode is taken account.
In addition, a real scalar field on a disconnected graph $G_S=C_3\cup C_3$ is
introduced. The `mass term' looks like
\be
f^2\phi_i\Delta(G_S)_{ij}\phi_{j}=
\frac{f^2}{2}\left[\sum_{i=1}^{3}\sum_{j=1}^{3}({\phi}_{i}-{\phi}_{j})^2+
\sum_{i=4}^{6}\sum_{j=4}^{6}({\phi}_{i}-{\phi}_{j})^2\right]\, ,
\ee
in the present case.

In this model, the one-loop vacuum energy density is calculated as
\bea
V_0=&+&\frac{3}{4\pi^2}\left(\frac{f}{6}\right)^4\sum_{q=1}^{\infty}
\frac{4\cos(6q\delta)-3}{q(q^2-1/36)(q^2-1/9)}\nn
&-&\frac{24}{\pi^2}\left(\frac{f}{6}\right)^4\sum_{q=1}^{\infty}
\frac{1}{q(q^2-1/9)(q^2-4/9)}\, ,
\eea
while the induced Newton constant becomes
\bea
\frac{1}{16\pi G_N}=&+&\frac{1}{48\pi^2}\left(\frac{f}{6}\right)^2\sum_{q=1}^{\infty}
\frac{2\cos(6q\delta)-3}{q(q^2-1/36)}\nn
&+&\frac{1}{6\pi^2}\left(\frac{f}{6}\right)^2\sum_{q=1}^{\infty}
\frac{1}{q(q^2-1/9)}\, .
\eea

%%%%%%%%%%%%%%%%%%%%%%%%%%%%%%%%%%%%%%%%%%%%%%%%%%%%%%%%%%%%%%%%%%%%%%
%%%FIGURES 2
%%%%%%%%%%%%%%%%%%%%%%%%%%%%%%%%%%%%%%%%%%%%%%%%%%%%%%%%%%%%%%%%%%%%%%
\begin{figure}[htb]
\centering
\mbox{\epsfbox{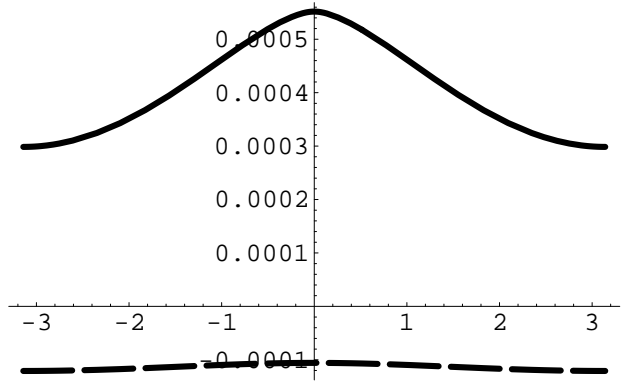}}\\
%\mbox{(a)}\\
%\mbox{\epsfbox{scalar2.eps}}\\
%\mbox{(b)}\\
\bigskip
\caption{%
$V_0$ (the dashed line) and $1/(16\pi G_N)$ (the bold line)
in the first model.}
\label{fig2}
\end{figure}
%%%%%%%%%%%%%%%%%%%%%%%%%%%%%%%%%%%%%%%%%%%%%%%%%%%%%%%%%%%%%%%%%%%%%%
%%%FIGURES 2
%%%%%%%%%%%%%%%%%%%%%%%%%%%%%%%%%%%%%%%%%%%%%%%%%%%%%%%%%%%%%%%%%%%%%%

The vacuum energy density and the Newton `constant' are plotted
against the link zero-mode in FIG.~\ref{fig2} for this model. We set $f=1$. The
horizontal axis indicates the value
$6\delta$. The bold line shows $1/(16\pi G_N)$ and the dashed line shows $V_0$.
We see that $G_N$ is finite and positive while $V_0$ is finite but negative.

\begin{figure}[htb]
\centering
\mbox{\epsfbox{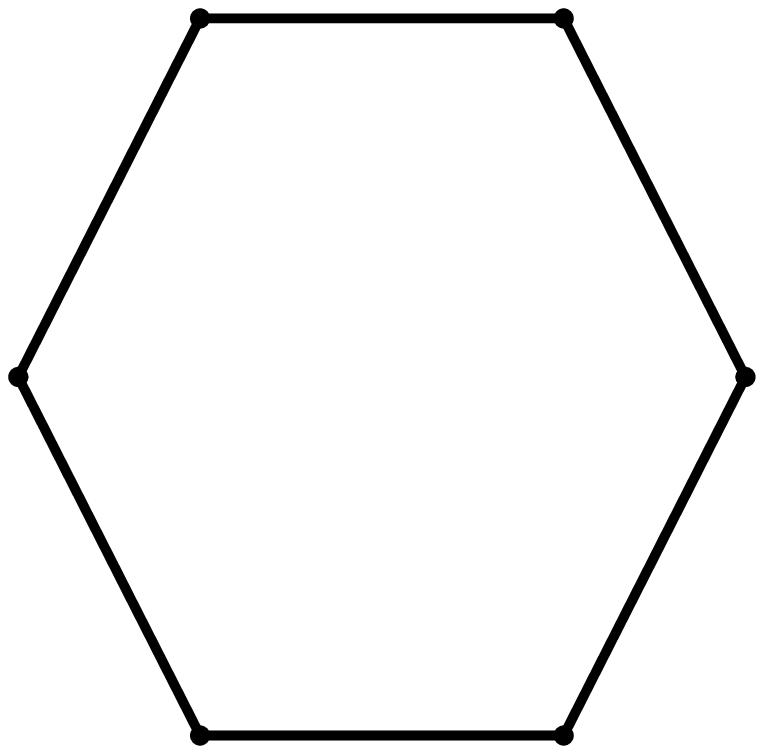}}\\
\mbox{($G_V$)}\\
\mbox{\epsfbox{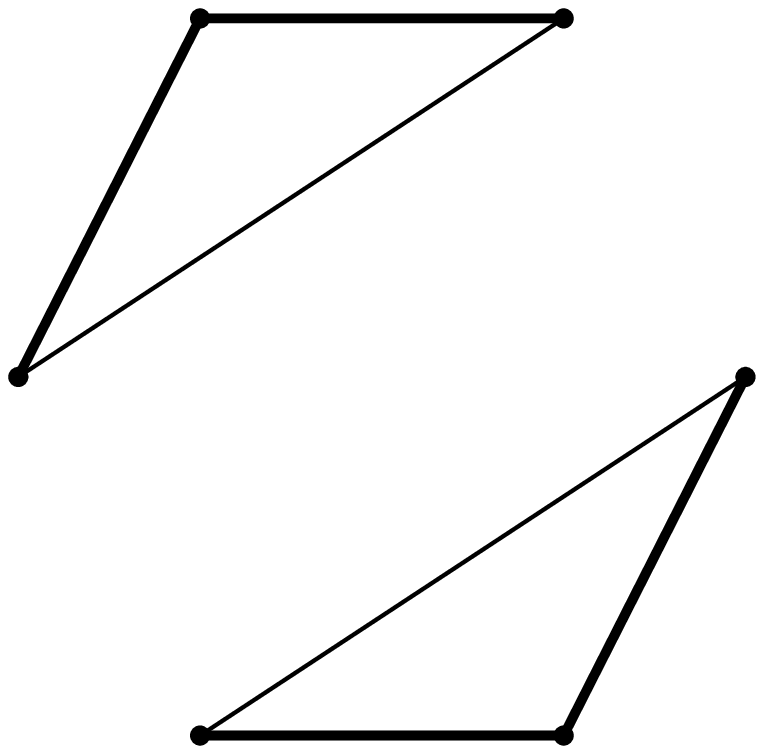}}\\
\mbox{($G_F$)}\\
\bigskip
\caption{$G_V$ and $G_F$ in the second minimal model. The bold edges indicate the
location of the link field (zero-mode).}
\label{fig3}
\end{figure}

Now we consider another model.
The vector field is considered on $G_V=C_6$ as in the first model, but
the Dirac field is on a disconnected graph $G_F=C_3\cup C_3$.
Although the coupling between the Dirac field and the link field is arbitrarily
selected, it is natural to assume that two edges of each $C_3$ is connected by the
link field. 
Then the phases from the link zero-mode are attached only to such edges. 
In FIG.~\ref{fig3}, we exhibit the coupling to the link fields.
The real scalar field is considered to be on a disconnected graph $C_3\cup C_3$
as before.

%%%%%%%%%%%%%%%%%%%%%%%%%%%%%%%%%%%%%%%%%%%%%%%%%%%%%%%%%%%%%%%%%%%%%%
%%%FIGURES 3
%%%%%%%%%%%%%%%%%%%%%%%%%%%%%%%%%%%%%%%%%%%%%%%%%%%%%%%%%%%%%%%%%%%%%%
%%%%%%%%%%%%%%%%%%%%%%%%%%%%%%%%%%%%%%%%%%%%%%%%%%%%%%%%%%%%%%%%%%%%%%
%%%FIGURES 3
%%%%%%%%%%%%%%%%%%%%%%%%%%%%%%%%%%%%%%%%%%%%%%%%%%%%%%%%%%%%%%%%%%%%%%

In this model, the one-loop vacuum energy density is calculated as
\bea
V_0=&-&\frac{9}{4\pi^2}\left(\frac{f}{6}\right)^4\sum_{q=1}^{\infty}
\frac{1}{q(q^2-1/36)(q^2-1/9)}\nn
&+&\frac{24}{\pi^2}\left(\frac{f}{6}\right)^4\sum_{q=1}^{\infty}
\frac{4\cos(2q\delta)-1}{q(q^2-1/9)(q^2-4/9)}\, ,
\eea
while the induced Newton constant is
\bea
\frac{1}{16\pi G_N}=&-&\frac{1}{16\pi^2}\left(\frac{f}{6}\right)^2\sum_{q=1}^{\infty}
\frac{1}{q(q^2-1/36)}\nn
&+&\frac{1}{6\pi^2}\left(\frac{f}{6}\right)^2\sum_{q=1}^{\infty}
\frac{2\cos(2q\delta)+1}{q(q^2-1/9)}\, .
\eea

%%%%%%%%%%%%%%%%%%%%%%%%%%%%%%%%%%%%%%%%%%%%%%%%%%%%%%%%%%%%%%%%%%%%%%
%%%FIGURES 4
%%%%%%%%%%%%%%%%%%%%%%%%%%%%%%%%%%%%%%%%%%%%%%%%%%%%%%%%%%%%%%%%%%%%%%
\begin{figure}[htb]
\centering
\mbox{\epsfbox{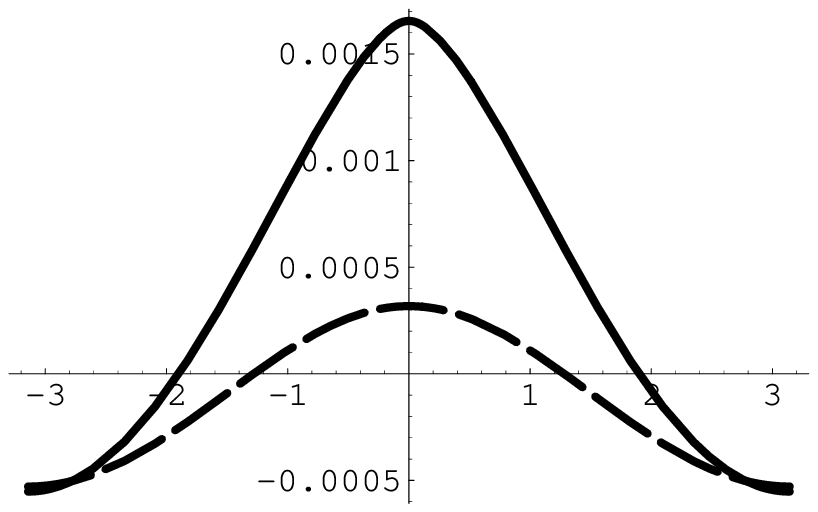}}\\
%\mbox{(a)}\\
%\mbox{\epsfbox{cp2.eps}}\\
%\mbox{(b)}\\
\bigskip
\caption{%
$V_0$ (the dashed line) and $1/(16\pi G_N)$ (the bold line) in
the second model.}
\label{fig4}
\end{figure}
%%%%%%%%%%%%%%%%%%%%%%%%%%%%%%%%%%%%%%%%%%%%%%%%%%%%%%%%%%%%%%%%%%%%%%
%%%FIGURES 4
%%%%%%%%%%%%%%%%%%%%%%%%%%%%%%%%%%%%%%%%%%%%%%%%%%%%%%%%%%%%%%%%%%%%%%

In FIG.~\ref{fig4}, we exhibit
the vacuum energy density and the Newton `constant' for the second model.
Each line shows the same as in the previous figure, but
the horizontal axis indicates the value
$2\delta$ in the present case. 
 We see that both $G_N$ and $V_0$ have definite sign.
Note that $G_N>0$ at the point satisfying $V_0=0$.

We should emphasize that these models are only minimal models.
Even if the matter content is fixed, the choice of graphs is arbitrary;
we can search the adequate graphs yielding the small cosmological constant.
But the possibility may be constrained by the requirement for the excited modes of
fields to decouple from the low energy phenomenology.
Roughly speaking, the total degree of a graph is larger the first excited mode is
more massive. Obviously, however, the number of the graphs with the common degree
matrix will decreases if the order of the graph is fixed.
Thus the possible graph may be restricted by the physical reason.
The detailed discussion including the general graph spectrum will be given in the
future work.

One more remark may be necessary. Besides we can incorporate the known tuning
mechanisms of the cosmological constant  so far, we can consider a new scenario taking
the link zero-mode field for a time-dependent field.
In the second minimal model shown above, the vacuum energy can be small if the
zero-mode field takes a certain value. Because this cannot be realized as a static
solution, we must consider dynamical evolution of the field in the similar possible
models, such as the mechanism suggested by Linde~\cite{Linde}.

%%%%%%%%%%%%%%%%%%%%%%%%%%%%%%%%%%%%%%%%%%%%%%%%%%%%%%%%%%%%%%%%%%%%%%
\section{summary and prospects}
\label{sec:5}
%%%%%%%%%%%%%%%%%%%%%%%%%%%%%%%%%%%%%%%%%%%%%%%%%%%%%%%%%%%%%%%%%%%%%%

In conclusion, we clarified the UV behavior of the one-loop effective lagrangian in
the theory on graphs. Using this knowledge,  we can build a theory space description
of the induced gravity.  More elaboration is in order for a small cosmological
constant.

We must study the following generalization.

We should consider more general case with the graph of $p\ne q$.
The gauge theory on such a graph will require
plaquette-like self-interaction of link fields.
In the same case, probably we cannot avoid the fermion (and chiral fermion zero-modes)
on edges. We may have more than the `copies' of the theory on vertices plus link
fields on edges. But we should remember that large arbitrariness is dangerous in
building  plausible models.

To consider the generalization of the Hosotani model,
we should investigate 
deconstruction of non-Abelian gauge theory.
We also need deconstruction of
adjoint matter fields or fields in other representations.

In this paper, we assumed only one mass scale ($f$).
If several mass scales are associated to edges, the corresponding graph has 
weighted edges.

We are interested also in the two-loop effective action.
We hope that the knowledge of graphs may be
useful to investigate higher-loop divergence as well as tree-level calculation of
reaction amplitude mediated by the excited modes.

\begin{acknowledgments}
We would like to thank Y.~Cho for his valuable comments
and for reading the manuscript.
\end{acknowledgments}

%\newpage

%%%%%%%%%%%%%%%%%%%%%%%%%%%%%%%%%%%%%%%%%%%%%%%%%%%%%%%%%%%%%%%%%%%%%%
%%%References
%%%%%%%%%%%%%%%%%%%%%%%%%%%%%%%%%%%%%%%%%%%%%%%%%%%%%%%%%%%%%%%%%%%%%%

%%%%%%%%%%%%%%%%%%%%%%%%%%%%%%%%%%%%%%%%%%%%%%%%%%%%%%%%%%%%%%%%%%%%%%

\begin{thebibliography}{99}

\bibitem{ACG} N.~Arkani-Hamed, A.~G.~Cohen and H.~Georgi,
Phys. Rev. Lett. {\bf 86} (2001) 4757; 
Phys. Lett. {\bf B513} (2001) 232. 

C.~T.~Hill, S.~Pokorski and J.~Wang,
Phys. Rev. {\bf D64} (2001) 105005. 

In early days, similar investigation has been done:
M.~B.~Halpern and W.~Siegel,
Phys. Rev. {\bf D11} (1975) 2967. 


\bibitem{HL} C.~T.~Hill and A.~K.~Leibovich,
%``Deconstructing 5-D QED'',
Phys. Rev. {\bf D66} (2002) 016006, 
{\tt hep-ph/0205057};
%``Natural Theories of Ultra-Low Mass PNGB's: Axions and Quitessence'',
Phys. Rev. {\bf D66} (2002) 075010, 
{\tt hep-ph/0205237}.

\bibitem{ma} 
A.~Dimakis and F.~M\"uller-Hoissen, J. Phys. {\bf A27} (1994) 3159, hep-th/9401149.

A.~Dimakis and F.~M\"uller-Hoissen, J. Math. Phys. {\bf 35} (1994) 6703,
hep-th/9404112.

A.~Dimakis, F.~M\"uller-Hoissen and F. Vanderseypen, J. Math. Phys. {\bf 36} (1995)
3771, hep-th/9408114.

M.~Requardt, J. Phys. {\bf A35} (2002) 759, math-ph/0108007.

Th.~Filk, Class. Q. Grav. {\bf 17} (2000) 4841, hep-th/0010126.

\bibitem{graphtheory} For an example,
R.~J.~Wilson,
{\it Introduction to Graph Theory} (4th Edition),
Longman, New York, 1997.

\bibitem{Lane} K.~Lane,
Phys. Rev. {\bf D65} (2002) 115001. 

\bibitem{fw} N.~Kan and K.~Shiraishi, in preparation.

\bibitem{specR}
A.~Mukherjee and I.~L.~Tabbash, Eur. Phys. J. C20 (2001) 193.

R.~Rabadan and G.~Shiu, JHEP 05 (2003) 045.

\bibitem{Chung} R.~K.~Chung,
{\it Spectral Graph Theory}, AMS, 1997. 

\bibitem{SS} J.~Scherk and J.~H.~Schwarz,
Phys. Lett. {\bf B82} (1979) 60; Nucl. Phys. {\bf B153} (1979) 61. 

\bibitem{Hosotani} Y.~Hosotani,
Phys. Lett. {\bf B126} (1983) 309.

D.~J.~Toms,
Phys. Lett. {\bf B126} (1983) 445. 

\bibitem{Sakharov} A.~A.~D.~Sakharov,
Soviet Physics Doklady {\bf 12} (1968) 1040,
reprinted in Gen. Rel. Grav. {\bf 32} (2000) 365.

\bibitem{Visser} M.~Visser, Mod. Phys. Lett. {\bf A17} (2002) 977.

\bibitem{KKIG}
D.~J.~Toms, Phys. Lett. {\bf B129} (1983) 31.

P.~Candelas and S.~Weinberg, Nucl. Phys. {\bf B237} (1984) 397.

I.~L.~Buchbinder, S.~D.~Odintsov and I.~L.~Shapiro, {\it Effective action in 
quantum gravity}, IOP Pub., Bristol, 1992.

\bibitem{Linde} A.~Linde, in {\it 300 years of gravitation},
edited by S.~W.~Hawking and W.~Israel, Cambridge University Press, 
Cambridge, 1987.

See also: A.~D.~Dolgov, in {\it The very early universe},
edited by G.~W.~Gibbons, S.~W.~Hawking and S.~T.~C.~Siklos, Cambridge University
Press,  Cambridge, 1983.








%
\end{thebibliography}
\end{document}